\documentclass[pra,twocolumn,superscriptaddress,showpacs,groupedaddress,floatfix]{revtex4-1}
\usepackage{amssymb}
\usepackage{graphicx}
\usepackage{subfigure}	 	
\usepackage[english]{babel}
\usepackage{float}
\usepackage{color}
\usepackage{bm}
\usepackage{dcolumn}
\usepackage{amsmath}
\usepackage{natbib}
\usepackage{epsfig}

\begin{document}

\title{Vortex solitons of the discrete Ginzburg-Landau Equation}
\author{C. Mej\'ia-Cort\'es}
\author{J.M. Soto-Crespo}
\affiliation{Instituto de \'Optica, C.S.I.C., Serrano 121, 28006 Madrid, Spain}

\author{Rodrigo A. Vicencio}
\author{Mario I. Molina}
\affiliation{Departamento de F\'isica, Facultad de Ciencias, Universidad de
Chile, Casilla 653, Santiago, Chile}
\affiliation{Center for Optics and Photonics, Universidad de Concepci\'on,
Casilla 4016, Concepci\'on, Chile}
\date{\today}

\begin{abstract}
We have found several families of vortex soliton solutions in two-dimensional
discrete dissipative systems governed by the cubic-quintic complex
Ginzburg-Landau equation. There are symmetric and asymmetric solutions, and some
of them have simultaneously two different topological charges. Their regions of
existence and stability are  determined. Additionally, we have analyzed the
relationship between dissipation and stability for a number of solutions. We
have obtained that dissipation favours the stability of the solutions.
\end{abstract}


\maketitle

\section{Introduction}

An optical vortex soliton is a self localized nonlinear wave, characterized for
having a point (`singularity') of zero intensity, and with a phase that twists
around that point, with a total phase accumulation of $2\pi S$ for a closed
circuit around the singularity~\cite{cvortopn}. The quantity $S$ is an integer
number known as the vorticity or topological charge of the solution. Optical
vortices can exist in  an infinite number of ways, as there is no limit to the
topological charge. This kind of waves looks attractive in future applications
for encoding  and storing information. A spatial vortex soliton is a specific
solution for a (2+1) dimensional nonlinear wave
equation~\cite{Desyatnikov2005291}. One of the most widely used equations of such
a type is the nonlinear Schr$\ddot{\text{o}}$dinger equation (NLSE); it
describes wave evolution in dispersive/diffractive continuous media with an
optical Kerr response, i.e. a refractive index that changes linearly with the
light intensity. When the system under consideration has a periodic structure,
e.g, a photonic crystal fiber, it  is necessary to add a periodic transversal
potential to complete the description, in the NLSE. The optical properties of a
nonlinear periodic structure can be analyzed in the framework of a set of
linearly coupled-mode equations  which, in solid-state physics is called the
{\it tight-binding} approximation, so that the description of the system can be
understood from a discrete point of view. The study of discrete systems has been
a hot topic in the last years due both to its broad impact in diverse branches
of science and its potential for technological
applications~\cite{campbell:43,Lederer20081,Flach20081,Vicencio:03}. Nonlinear
optical systems allow us to observe several self-localized discrete structures in
both spatial and temporal domains. 

Unlike conservative systems, self-localized structures in systems far from
equilibrium, are dynamical solutions that exchange energy with an external
source (open systems). These solutions are called dissipative
solitons~\cite{akhm05}. In Schr$\ddot{\text{o}}$dinger models, gain and loss are
completely neglected and the dynamical equilibrium is reached by means of a
balance between the Kerr effect and dispersion/diffraction. For
dissipative systems, there must also exist an additional balance between gain
and losses, turning the equilibrium into a many-sided
process~\cite{SotoCrespo2001283}.  The Ginzburg-Landau equation is -somehow- a
universal model where dissipative solitons are their most interesting solutions.
This model appears in many branches of science like, for example, nonlinear
optics, Bose-Einstein condensates, chemical reactions, super-conductivity and
many others~\cite{akhm0508}.

Nonlinear self-localized structures in optical lattices, usually referred to as
discrete solitons, have been predicted and observed for one- and two-dimensional
arrays~\cite{PhysRevLett.81.3383,fleis}. The existence of discrete vortex
solitons in conservative systems have been reported on several
works~~\cite{PhysRevE.64.026601,PhysRevLett.92.123903,PhysRevLett.102.224102}.
For the continuous case, dissipative vortex soliton families have been found to
be stable for a wide interval of
$S$-values~\cite{Soto-Crespo:09,PhysRevLett.105.213901}. Symmetric stable
vortices have also been predicted in continuous dissipative systems with a
periodic linear modulation~\cite{PhysRevA.80.033835}. In this work, we deal with
discrete vortex solitons in dissipative 2D lattices governed by a discrete
version of the Ginzburg-Landau equation. We have found different families of
these self-localized solutions. We studied their stability and found stable
vortex families for $S=3$ (symmetric) and $S=2$ (asymmetric) topological charges
for the same set of equation parameters. In addition, we found another symmetric
solution in which two topological charges ($S=2$ and $S=6$) coexist. Finally, we
show how an increase in dissipation increases the stability regions for the same
``swirl-vortex" soliton analyzed in the recent work~\cite{PhysRevA.82.063818}

The paper is organized as follows. In Section \ref{mod} we portray the model
that we are going to use in the rest of the paper. Sections \ref{res1} and
\ref{res2} describe the new families of solutions we obtain, and in Section
\ref{com} we compare the results of our dissipative model with the conservative
cubic case (Schr\"odinger limit). Finally, section \ref{con} summarizes our main
results and conclusions.

\section{Model}
\label{mod}
\subsection{The cubic quintic Ginzburg-Landau equation}

Beam propagation in 2D dissipative waveguide lattices can be modeled by the following
equation:
\begin{eqnarray}
\nonumber
&i\dot\psi_{m,n}+\hat C\psi_{m,n}+|\psi_{m,n}|^{2}\psi_{m,n}+\nu|\psi_{m,n}|^{4}\psi_{m,n}=\\
&i\delta\psi_{m,n}+i\varepsilon|\psi_{m,n}|^{2}\psi_{m,n}+i\mu|\psi_{m,n}|^{4}\psi_{m,n}\ .
\label{dgl2}
\end{eqnarray}
Eq.(\ref{dgl2}) represents a physical model for open systems that exchange
energy with external sources and it is called $(2+1)D$ discrete complex
cubic-quintic Ginzburg-Landau (CQGL) equation. $\psi_{m,n}$ is the complex field
amplitude at the $(m,n)$ lattice site and $\dot\psi_{m,n}$ denotes its
first derivative with respect to the propagation coordinate $z$. The set

\begin{displaymath}
\{m=-M,...,M\}\times\{n=-N,...,N\},
\end{displaymath}
defines the array, $2M+1$ and $2N+1$ being the number of sites in the horizontal
and vertical directions (in all our simulations $M=N=8$). The {\it tight
binding} approximation establishes that the fields propagating in each waveguide
interact linearly only with nearest-neighbor fields through their evanescent
tails. This interaction is described by the discrete diffraction operator
\begin{equation*}
\hat C\psi_{m,n}=C(\psi_{m+1,n}+\psi_{m-1,n}+\psi_{m,n+1}+\psi_{m,n-1}),
\end{equation*}
where $C$ is a complex parameter. Its real part indicates the strength  of the
coupling between adjacent sites and its imaginary part denotes the gain or loss
originated by this coupling. The nonlinear higher order Kerr term is represented
by $\nu$ while $\varepsilon>0$ and $\mu < 0$ are the coefficients for cubic gain
and quintic losses, respectively. Linear losses are accounted for  a negative
$\delta$. 

In contrast to the conservative discrete nonlinear Schr$\ddot{\text{o}}$dinger
(DNLS) equation, the optical power, defined as
\begin{equation}
Q(z)=\sum_{m,n=-M,-N}^{M,N}\psi_{m,n}(z)\psi_{m,n}^{*}(z)
\end{equation}
is not a conserved quantity in the present model. However, for a self-localized
solution, the power and its evolution will be the main quantity that we will
monitor in order to identify different families of stationary and stable solutions.

We look for stationary solutions of Eq.(\ref{dgl2}) of the form
$\psi_{m,n}(z)=\phi_{m,n}\exp(i\lambda z)$ where $\phi_{m,n}$ are complex
numbers and $\lambda$ is real; also we are interested in that the phase of
solutions changes azimuthally an integer number ($S$) of $2\pi$ in a
closed-circuit. In such a case, the self-localized solution is called a discrete
vortex soliton~\cite{Pelinovsky200520} with vorticity $S$. By inserting the
previous {\em ansatz} into model (\ref{dgl2}) we obtain the following set of
$(2M+1)\times(2N+1)$ algebraic coupled complex equations:
\begin{eqnarray}
\nonumber
&-\lambda\phi_{m,n}+\hat C\phi_{m,n}+|\phi_{m,n}|^{2}\phi_{m,n}+\nu|\phi_{m,n}|^{4}\phi_{m,n}=\\
&i\delta\phi_{m,n}+i\varepsilon|\phi_{m,n}|^{2}\phi_{m,n}+i\mu|\phi_{m,n}|^{4}\phi_{m,n}\ .
\label{adgl2}
\end{eqnarray}
We have solved equations (\ref{adgl2}) using a multi-dimensional Newton-Raphson
iterative algorithm. The method requires an initial guess, and we have found
that usually converges rapidly by starting with a high-localized profile seed
that can be constructed by a procedure similar  to the one described
in~\cite{PhysRevA.82.063818}.

\subsection{Linear stability analysis}

Small perturbations around the stationary solution can grow exponentially,
leading to the destruction of the vortex soliton. A linear stability analysis
provides us the means for establishing which solutions are stable. Let us to
introduce a small perturbation, $\tilde\phi$, to the localized stationary solution 

\begin{equation}
\psi_{m,n}=[\phi_{m,n}+\tilde\phi_{m,n}(z)]e^{i\lambda z},\hspace{1cm}\tilde\phi_{m,n}\in\mathbb{C}
\label{Eq:pert},
\end{equation}
then, after replacing Eq.(\ref{Eq:pert}) into Eq.(\ref{dgl2}) 
and after linearizing with
respect to $\tilde\phi$, we obtain:

\begin{eqnarray}
\nonumber
&\dot{\tilde\phi}_{m,n}+\hat C\tilde\phi_{m,n}-i\delta\tilde\phi_{m,n}+\\
\nonumber
& [2(1-\varepsilon)|\phi_{m,n}|^{2}+3(\nu-\mu)|\phi_{m,n}|^{4}-\lambda]\tilde\phi_{m,n}+\\
&[(1-\varepsilon)\phi_{m,n}^{2}+2(\nu-\mu)|\phi_{m,n}|^{2}\phi_{m,n}^{2}]\tilde\phi^{*}_{m,n}=0.\
\label{dgl2p}
\end{eqnarray}

The solutions for the above homogeneous linear system can be written as
\begin{equation}
\tilde\phi_{m,n}(z)=C^{1}_{m,n}\exp{[\gamma_{m,n}z]}+C^{2}_{m,n}\exp{[\gamma_{m,n}^{*} z]}
\label{solpert},
\end{equation}
being $C^{1,2}$ integration constants and $\gamma_{m,n}$ the discrete spectrum
of the $eigensystem$ associated with (\ref{dgl2p}). The solutions are unstable
if at least one eigenvalue has positive real part, that is, if
$\textsf{max}\{\text{Re}(\gamma_{m,n})\}>0$. Hereafter, we will plot stable
(unstable) solutions using solid (dashed) lines.

 
\section{Symmetric and asymmetric solutions}
\label{res1}

Eq.(\ref{dgl2}) has a five-dimensional parameter space, namely
$C,\delta,\varepsilon,\mu,\nu$. In order to look for any stationary solution,
first, we need to choose a fixed set of values for these parameters, and then an
initial condition. By starting from a guess with eight peaks surrounding the
central site - the first discrete contour of the lattice around of
$(m,n)=(0,0)$- with a topological charge $S=3$ sampled on this path, the
iterative algorithm rapidly converges to a stationary structure with the same
features of the initial guess. Once we found a stationary solution with the
desired properties, we use it as initial condition to find the corresponding
solution for a slightly different set of equation parameters. We usually just
change one of them. Therefore, for the dissipative case we construct families of
solutions by fixing four parameters and varying the fifth one, usually the cubic
gain parameter, $\varepsilon$.
\begin{figure}
\centering
\epsfig{file=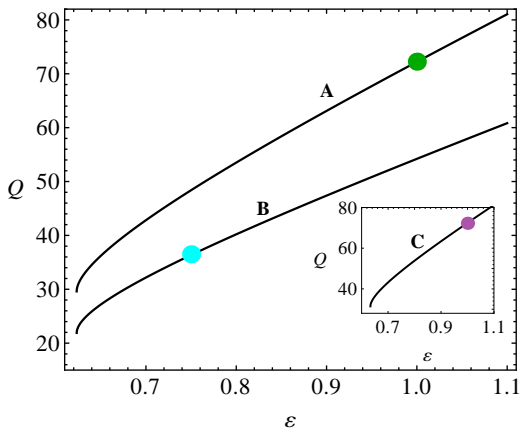,width=0.8\linewidth,clip=}
\caption{(Color online) $Q$ versus $\varepsilon$ diagram for  
families {\bf A} and {\bf B} of discrete dissipative vortex solitons. 
Inset: $Q$ vs. $\varepsilon$ diagram for family {\bf C}.  
(CQGL equation parameters: $C=0.8$, $\delta=-0.9$, $\mu=-0.1$, $\nu=0.1$).}
\label{fig1}
\end{figure}
Using this procedure, we have constructed the {\bf A} family (displayed as the curve $Q$
versus $\varepsilon$ in Fig.\ref{fig1}). We started from a highly
localized solution and we slowly decreased the nonlinear gain, observing that the
solution became gradually more and more extended as $\varepsilon$ (and $Q$)
decreased. The saddle-node point for this family is reached at
$\varepsilon\approx 0.62$.

\begin{figure}[h]
\centering
\begin{tabular}{cc}
\epsfig{file=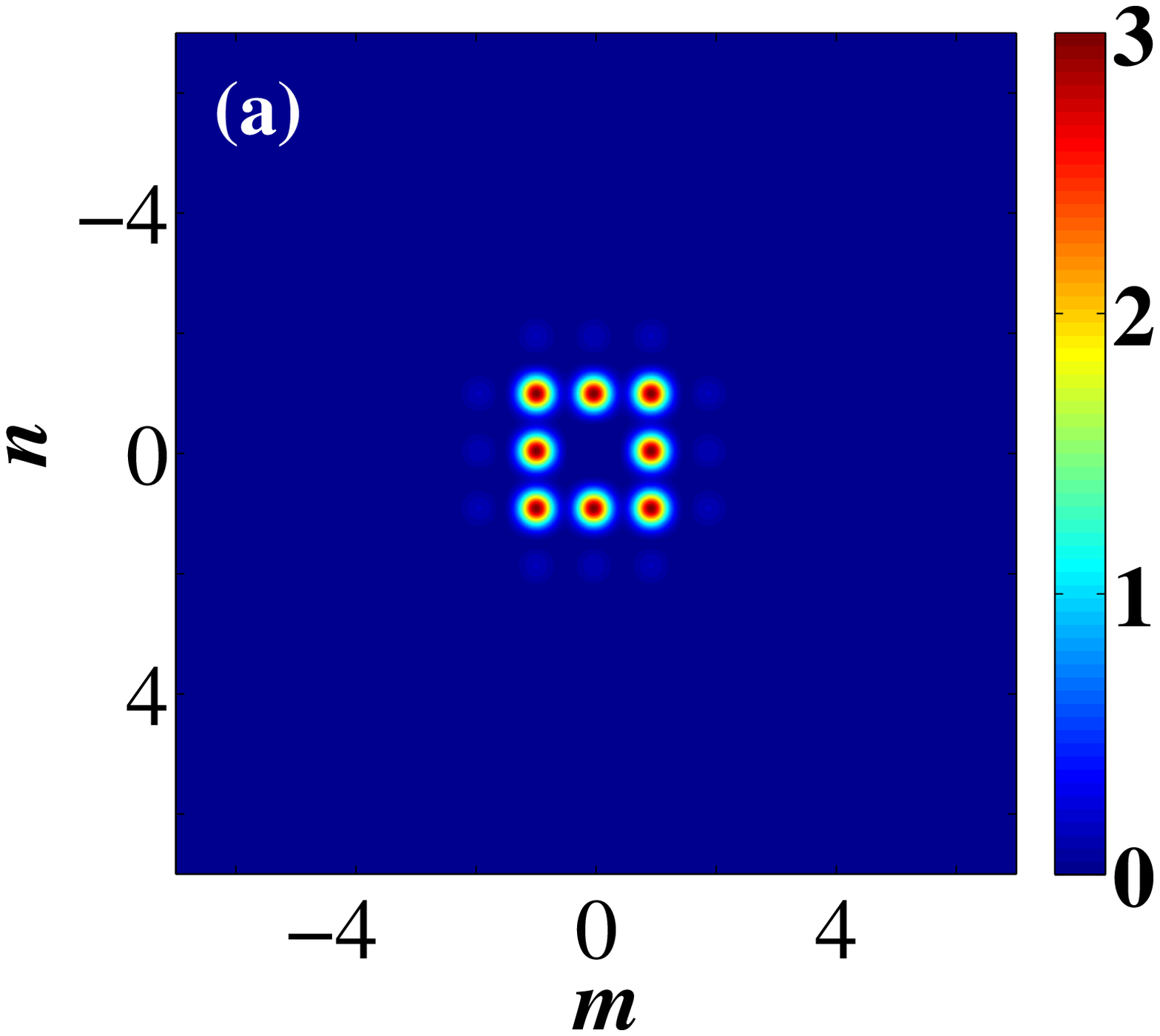,width=0.48\linewidth,clip=}&
\epsfig{file=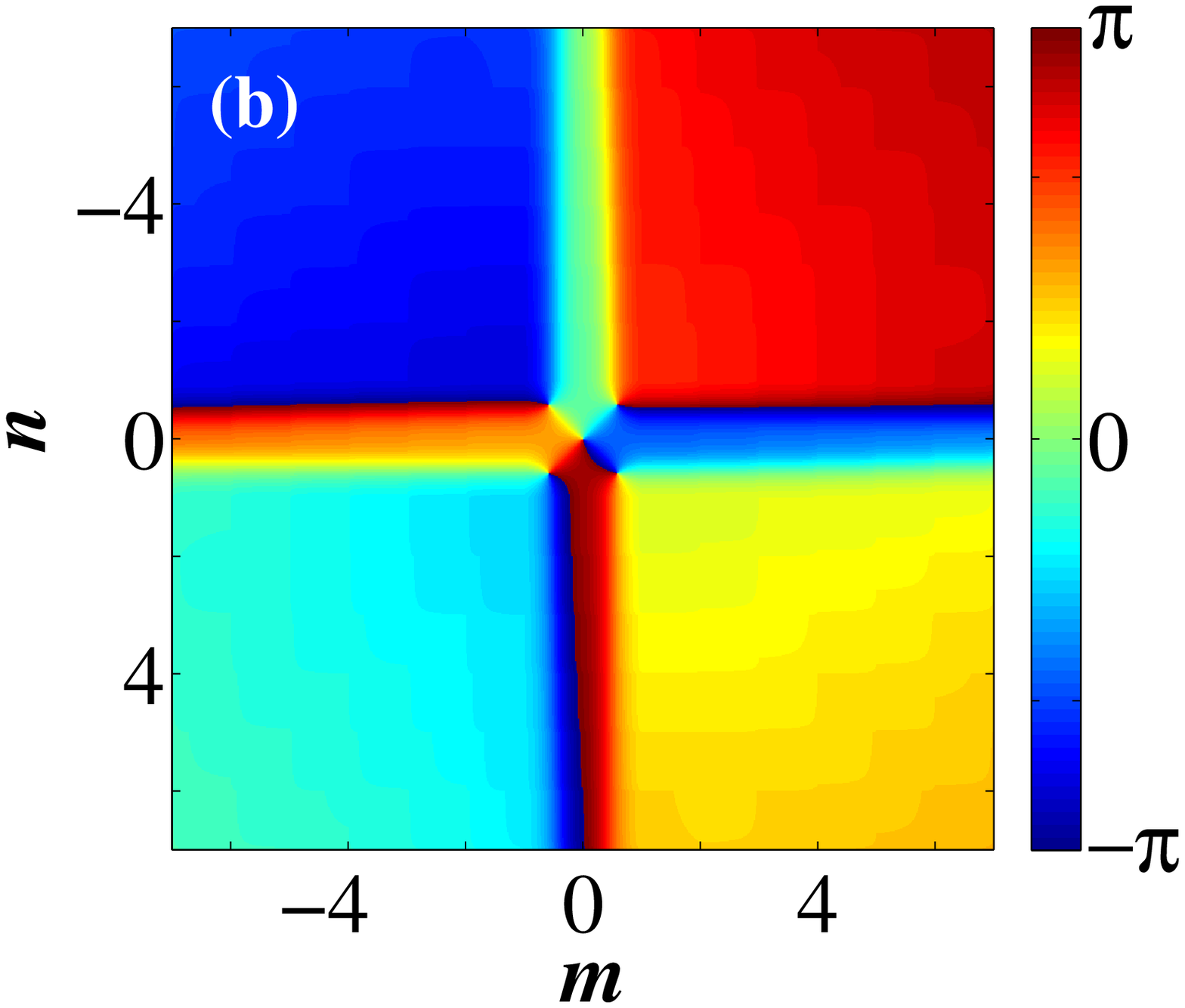,width=0.5\linewidth,clip=}
\end{tabular}
\caption{(Color online) Color map plots for the eight peaks stable vortex
 solution with $S=3$, marked with a green circle on the {\bf A} family branch 
 in Fig.\ref{dgl2}. (a) Amplitude profile. (b) Phase profile.}
\label{fig2}
\end{figure}

Fig.\ref{fig2} shows the amplitude and phase profiles corresponding to the
solution marked with a green solid circle on the {\bf A} family in
Fig.\ref{fig1}). From the amplitude profile, Fig.\ref{fig2}(a), we can see how
the stationary solution maintains the eight excited peaks of the initial seed.
Besides, we can see some energy in the tails, i.e. on the second discrete
contour. On the other hand, the phase profile, Fig.\ref{fig2}(b), clearly shows
a topological charge $S=3$.

\begin{figure}
\centering
\begin{tabular}{cc}
\epsfig{file=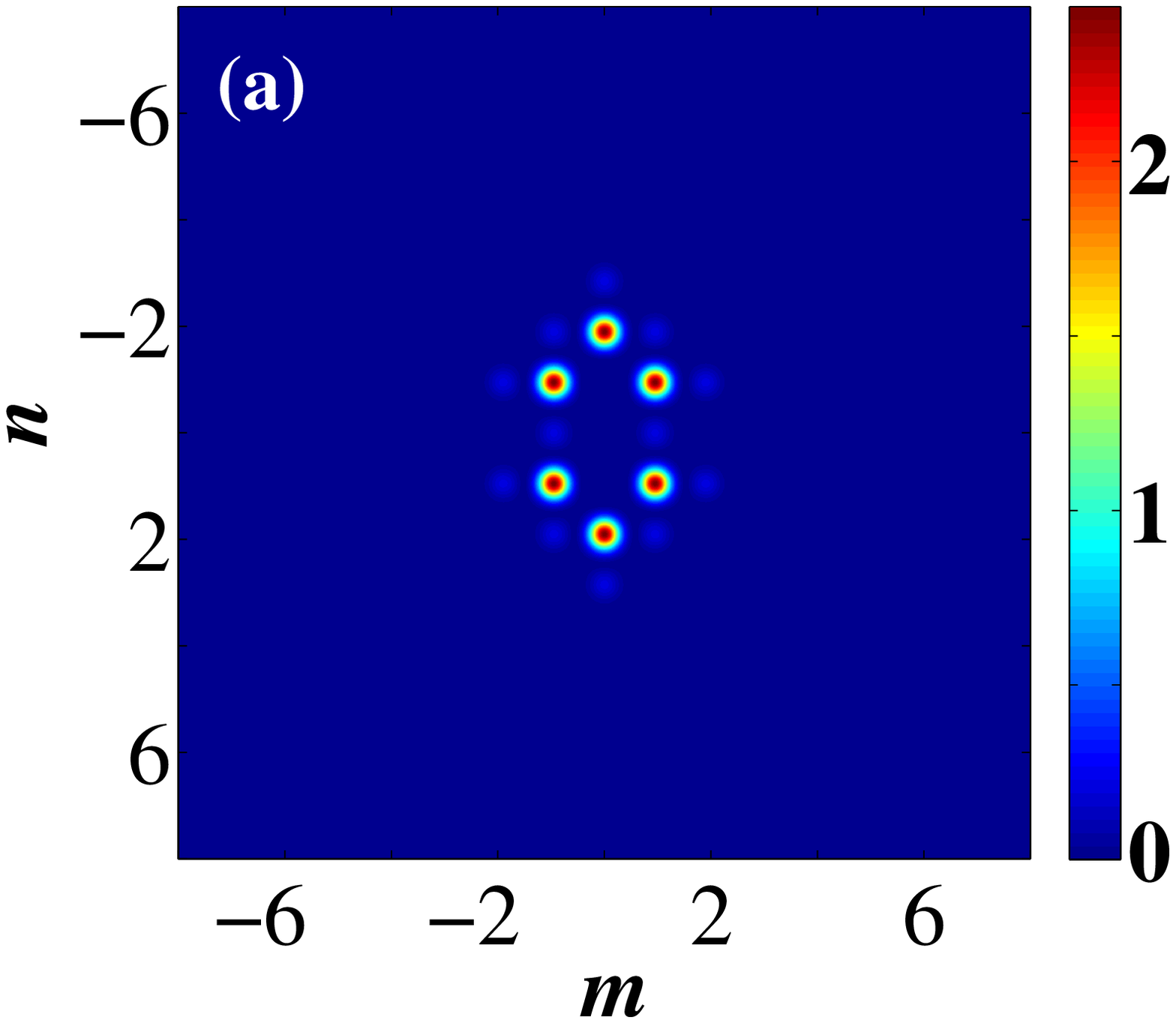,width=0.51\linewidth,clip=}&
\epsfig{file=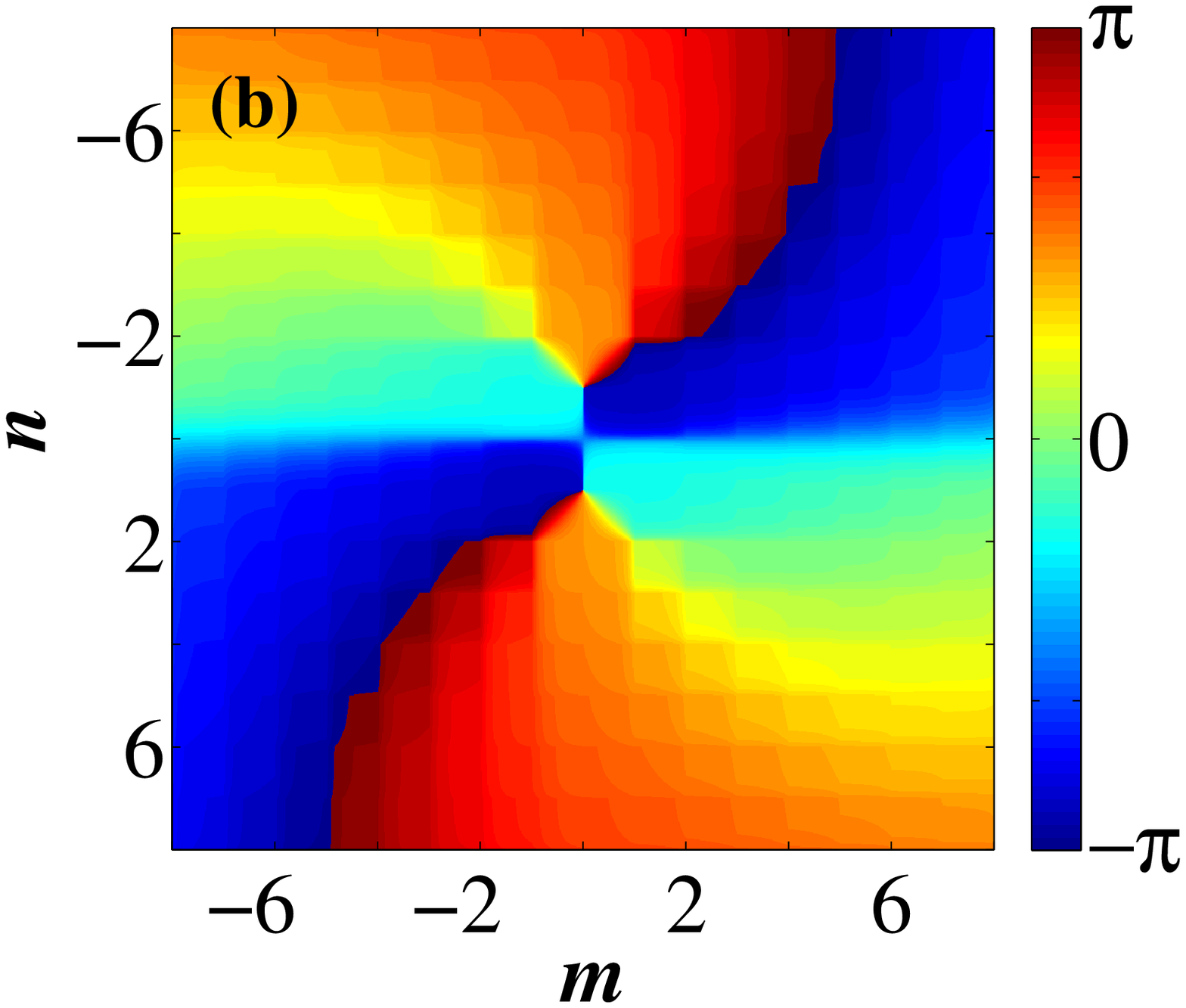,width=0.5\linewidth,clip=}
\end{tabular}
\caption{(Color online) Color map plots for the six peaks stable vortex solution with $S=2$
marked with a cyan circle on the {\bf A} family branch in Fig.\ref{dgl2}.
(a) Amplitude profile. (b) Phase profile.}
\label{fig3}
\end{figure}

A similar procedure has been done to construct another family, labeled {\bf B}
(See Fig.\ref{fig1}). This family consists of  asymmetric stationary solutions
characterized for having six peaks displayed on the corners of an elongated
hexagon in the $n$-axis direction of the lattice. This spatial configuration
possesses a topological charge $S=2$. Typical amplitude and phase profiles for
this kind of solution are  shown as color maps in Fig.\ref{fig3}. In the
conservative case, four peaks structures have been reported to be
stable~\cite{Pelinovsky200520} for $S=1$ and unstable for $S=2$; on the other
hand, for hexagonal lattices, six peak structures are
stable~\cite{PhysRevA.79.043821} for $S=2$ and unstable for $S=1$. In continuous
systems assymetric four peaks structures has been found stable for
$S=1$~\cite{PhysRevLett.93.063901}.

The families {\bf A} and {\bf B} of stationary vortex solutions coexist for the
same set of parameters of the discrete Ginzburg-Landau equation.  Other families
of solutions exist too for the same set of parameters. The inset shows a
different family (the {\bf C}-one), whose $Q$ vs $\varepsilon$ diagram almost
coincides with the {\bf A}-family one, in  spite of  being quite different type
of solutions. We will describe  these {\bf C}-solutions later in Section
\ref{com}.

\section{``Two charges" vortex soliton}
\label{res2}

\begin{figure}
\centering
\begin{tabular}{cc}
\epsfig{file=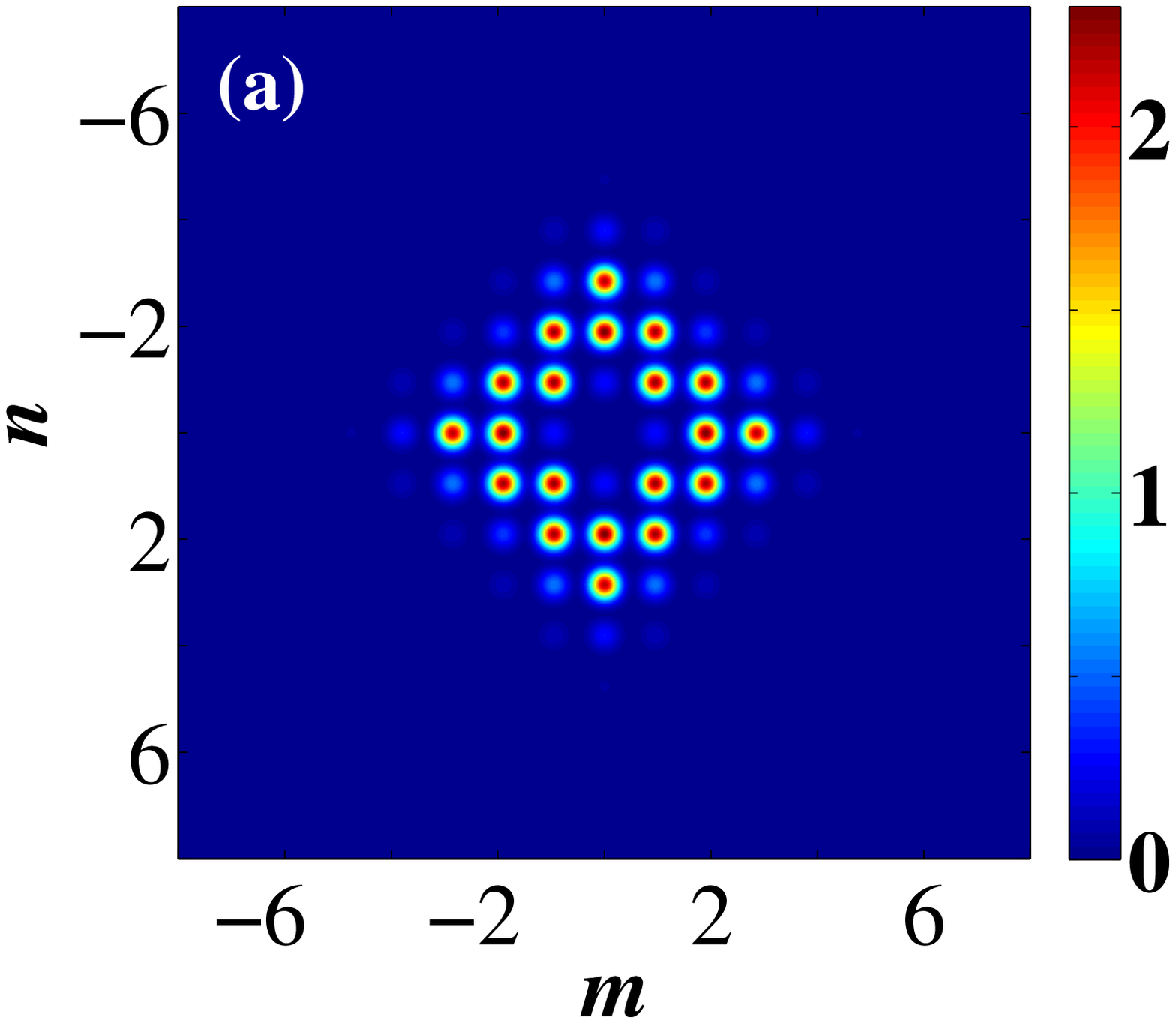,width=0.51\linewidth,clip=}&
\epsfig{file=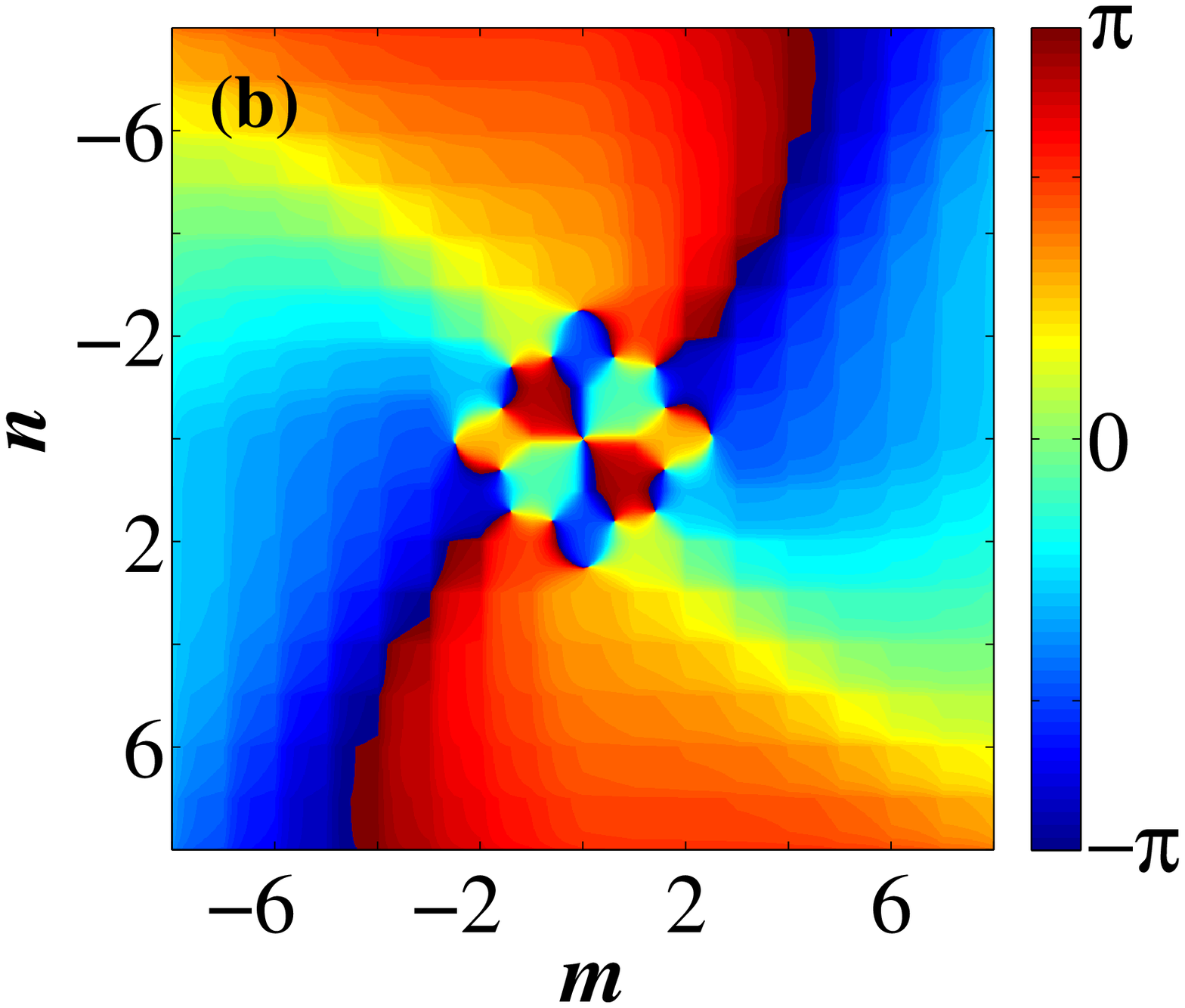,width=0.5\linewidth,clip=}
\end{tabular}
\caption{(Color online) Color map plots for the twenty peaks stable two charges 
($S=2$ and $S=6$) vortex solution, marked with a green circle on the {\bf D} family 
in Fig.\ref{fig5}.
(a) Amplitude profile. (b) Phase profile.}
\label{fig4}
\end{figure}

\begin{figure}
\centering
\epsfig{file=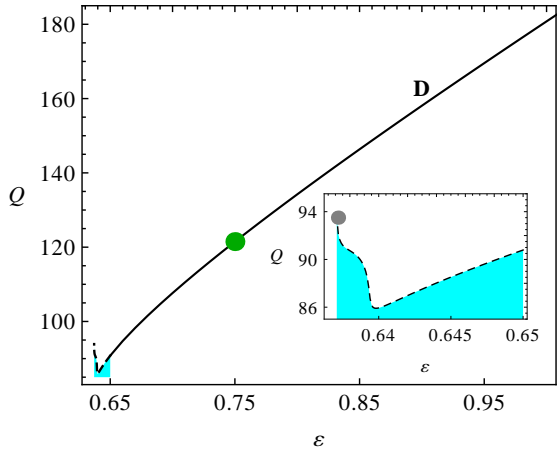,width=0.85\linewidth,clip=}
\caption{(Color online) $Q$ versus $\varepsilon$ diagram for ``two charges" ($S=2$
and $S=6$) discrete vortex solitons. Continuous and dashed lines correspond to
stable and unstable solutions, respectively. The green circle on the {\bf D} family
corresponds to the profiles shown in Fig.\ref{fig4}.} \label{fig5}
\end{figure}

Now, we show one example where two topological charges coexist in the same
solution. Let us start with a guess solution consisting of twenty peaks,
spatially distributed like a rhombus,  and with  a topological charge $S=2$.
Using it as the starting point for the Newton-Raphson algorithm, we find a
stationary solution that looks like the one shown in Fig.\ref{fig4}(a,b).
This solution belongs to the family displayed in Fig.\ref{fig5} labeled with
{\bf D}; it was constructed following the same procedure described in the previous
section. Unlike the previous families, the {\bf D} family does not reach the saddle
node point {\em via} a monotonic decreasing of its power; rather, it passes
through a minimum value ($\varepsilon\approx 0.64$), then, the power grows and
finally, the saddle node point is reached (See inset in Fig.\ref{fig5}).

\begin{figure}
\centering
\epsfig{file=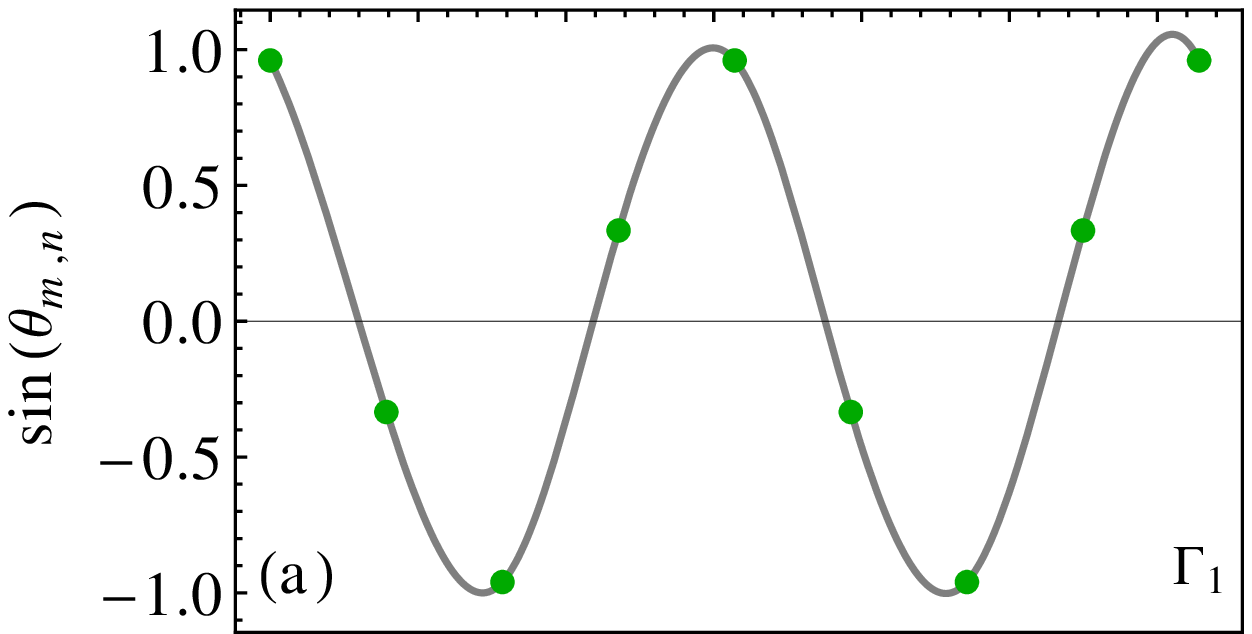,width=0.75\linewidth,clip=}\\
\epsfig{file=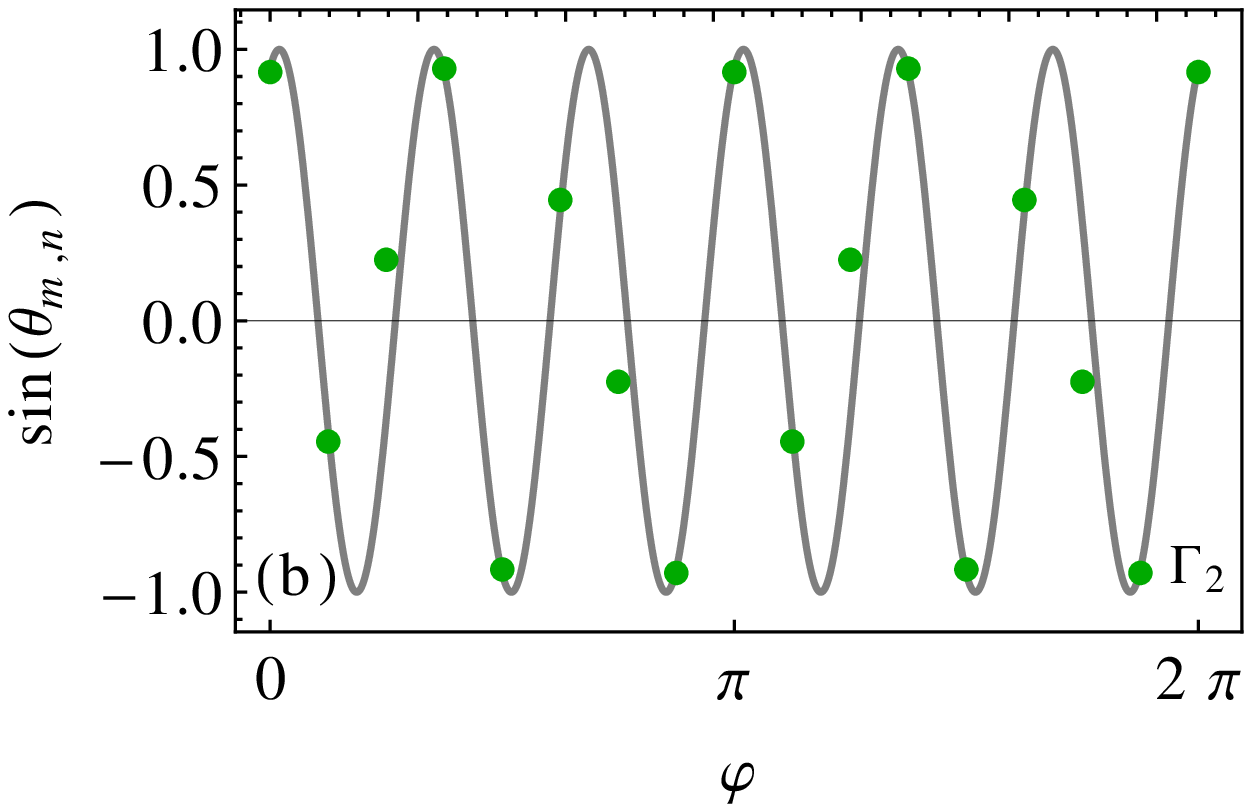,width=0.75\linewidth,clip=}
\caption{(Color online) $\sin(\theta_{m,n})$ versus $\varphi$ (azimuthal
angle for the lattice) diagram for the first (a) and second (b) 
discrete contour for the vortex soliton, marked with a green
dot on the {\bf D} family in Fig.\ref{fig5}}
\label{fig6}
\end{figure}

The solutions of this family present a very interesting property related to its
topological charge. The first square contour, $\Gamma_{1}$, -the innermost
discrete square trajectory on the plane $(m,n)$- in Fig.\ref{fig4}(b) shows that
the vorticity has a value $S=2$. For the second contour $\Gamma_{2}$ we observe
that the topological charge has changed to $S=6$. Looking at the remaining
contours, we note that the topological charge returns to $S=2$, so, we can talk
about a transition of the effective vorticity from $S=2\rightarrow
S=6\rightarrow S=2$, as we move farther from the center. For this reason, we can
say that the stable solutions of this family possess two topological charges.
For the sake of clarity, we plot $\sin(\theta_{m,n})$ vs $\varphi$, the
azimuthal angle for the lattice, for the first and second discrete contours.
From Fig.\ref{fig6}(a) we can see that the data (green points) are perfectly
fitted by the sinusoidal function (gray line) with two periods ($S=2$) along the
first contour, and for the second contour we have six periods ($S=6$) as shown
in Fig.\ref{fig6}(b). This is somehow a proof of the different topological
charges contained in the solution, and it also proves that the discrete vortex
is a well defined structure.

\begin{figure}
\centering
\begin{tabular}{cc}
\epsfig{file=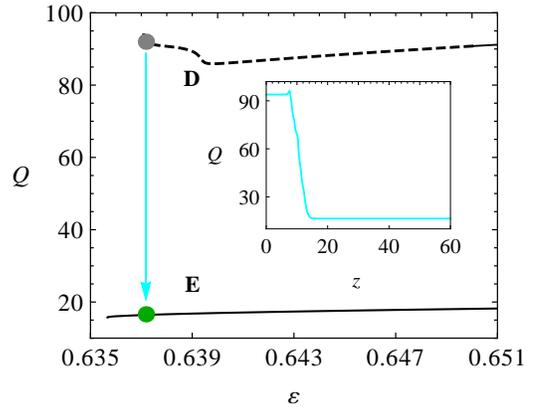,width=0.8\linewidth,clip=}
\end{tabular}
\caption{(Color online) $Q$ versus $\varepsilon$ diagram showing
the transition from the unstable
solution marked with the gray circle to the stable solution marked with the green circle;
the inset shows the power evolution for this transition.} 
\label{fig7}
\end{figure}

\begin{figure}
\centering
\begin{tabular}{cc}
\epsfig{file=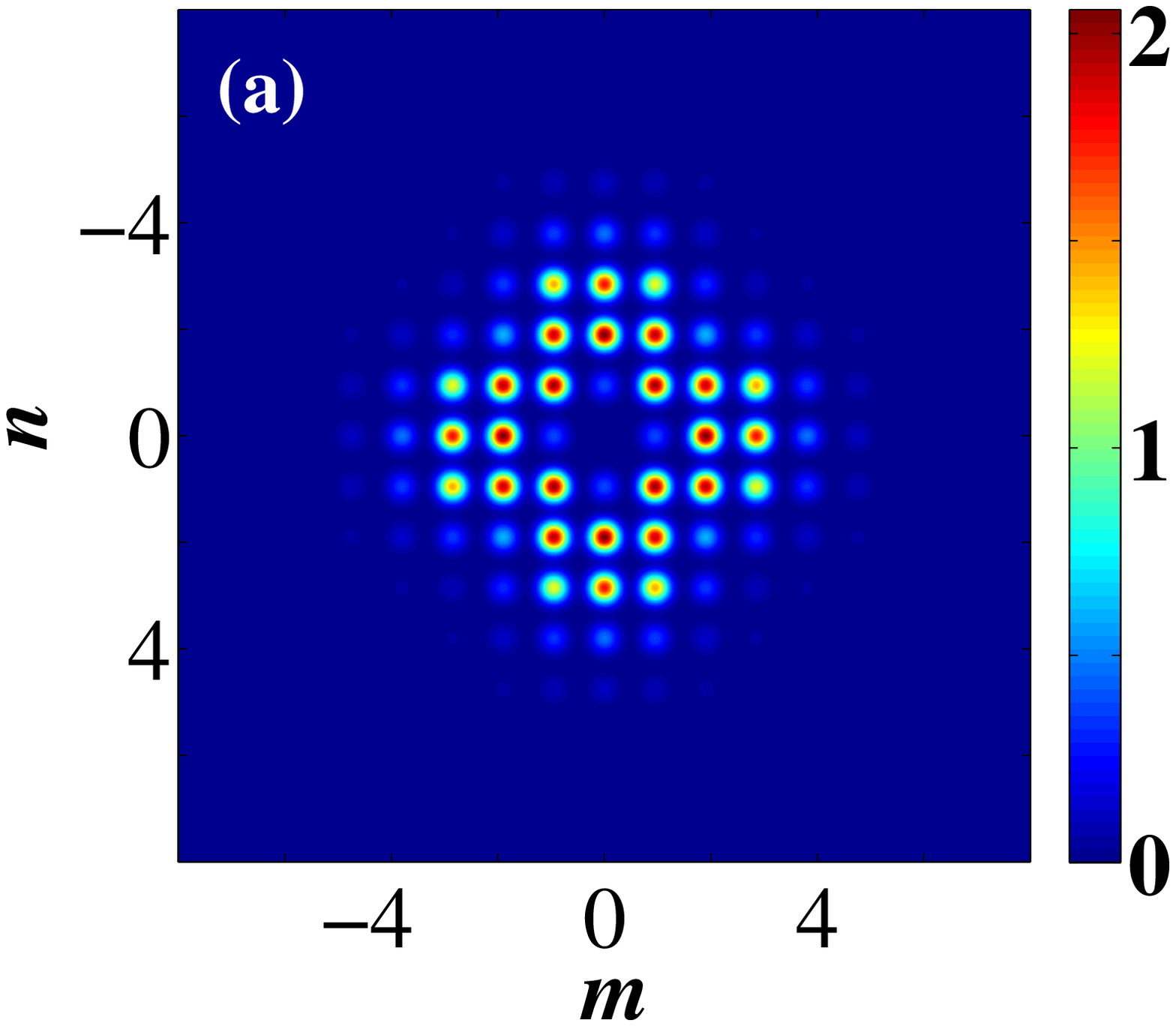,width=0.46\linewidth,clip=}&
\epsfig{file=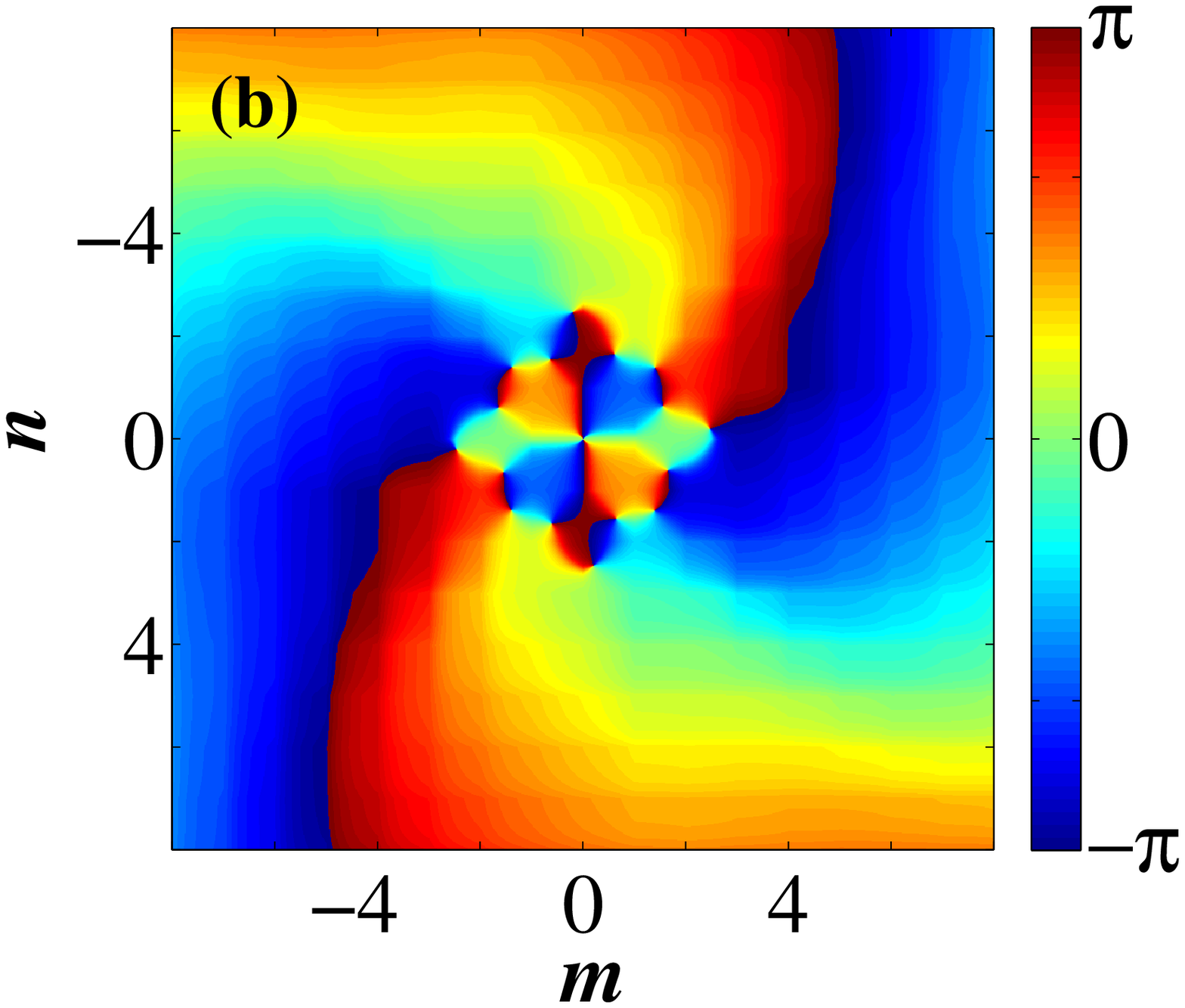,width=0.45\linewidth,clip=}\\
\epsfig{file=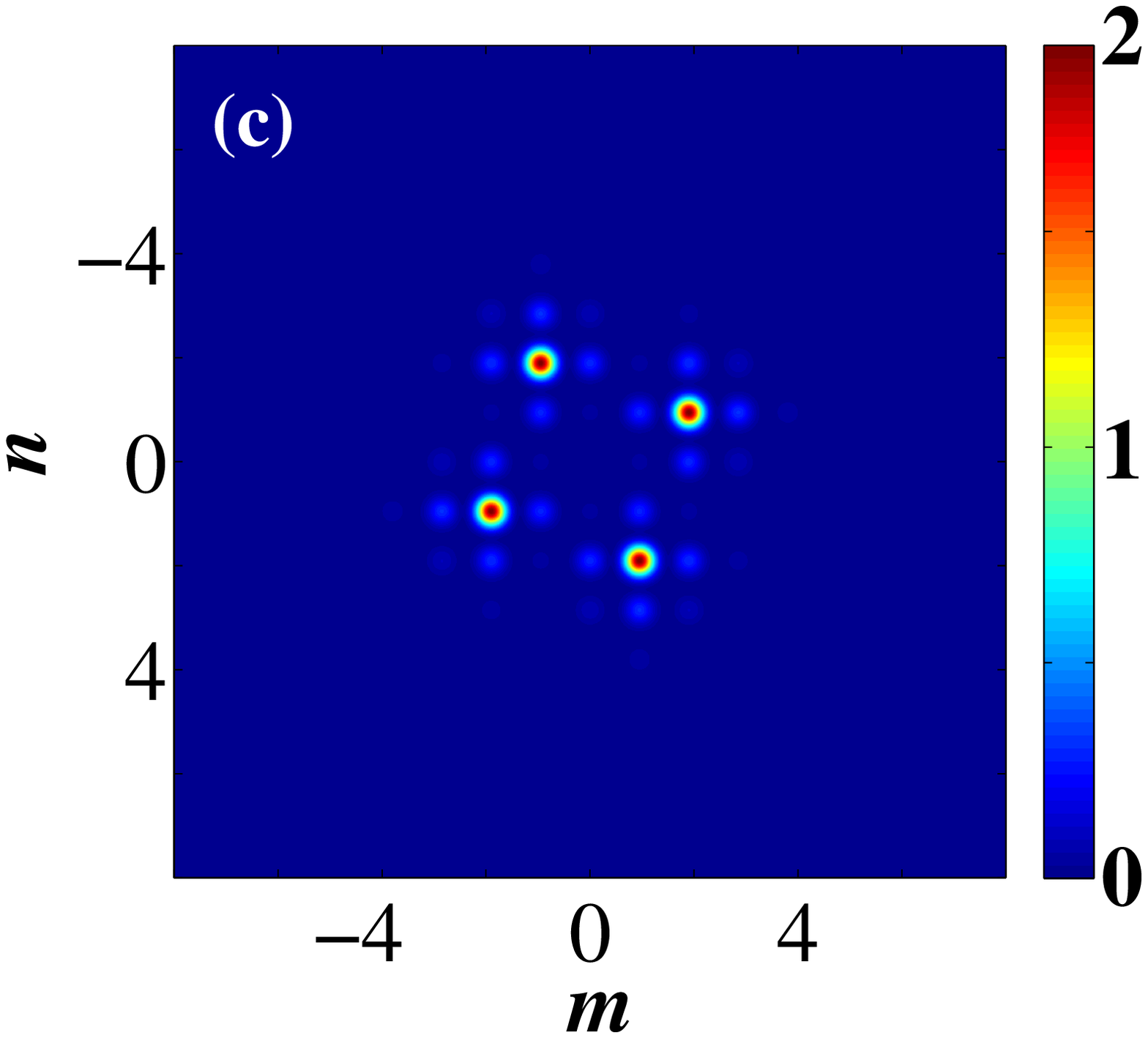,width=0.46\linewidth,clip=}&
\epsfig{file=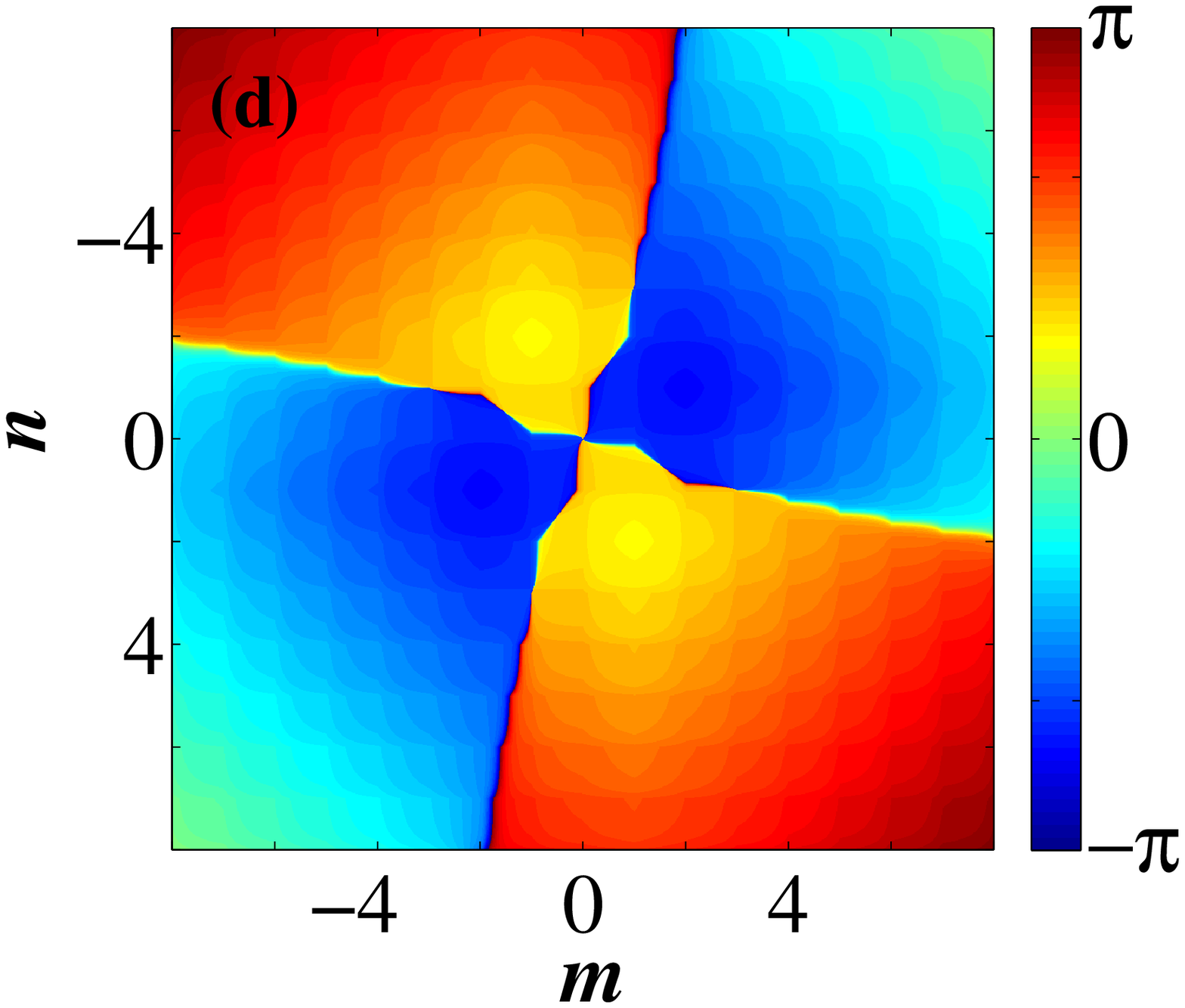,width=0.45\linewidth,clip=}
\end{tabular}
\caption{(Color online) Color map plots for discrete dissipative solitons. (a)
Amplitude profile and (b) phase profile for the unstable twenty peaks vortex
solution localized on the {\bf D} family at the gray circle in the Fig.\ref{fig7}.
(c) Amplitude profile and (d) phase profile for the four peaks soliton solution
localized on the {\bf E} family at the green circle in the Fig.\ref{fig7}}
\label{fig8}
\end{figure}

Fig.\ref{fig5} also shows that the {\bf D} family has one large stable region
and another small region, magnified in the inset, where the solutions are unstable.
These unstable structures decay on propagation to another kind of stable solutions
having less energy, and different amplitude profile with only four peaks. In
particular, Fig.\ref{fig7} illustrates how the unstable solution marked with
a gray point (the saddle-node point for the {\bf D} family in the Fig.\ref{fig4})
decays, by means of a radiative process shown in the inset, to the  stable solution
marked with a green circle on the {\bf E} family. The amplitude and phase profiles
showed in Figs.\ref{fig8}(a-b) and Figs.\ref{fig8}(c-d) correspond to the
unstable and stable solutions marked with gray and green circles in
Fig.\ref{fig7}.  

\section{Dissipation and stability}
\label{com}

In this section, we are interested in analyzing  how the stability of the solutions is
affected when our model slowly goes to the Schr$\ddot{\text{o}}$dinger limit,
i.e when the value of the parameters in the CQGL equation (\ref{dgl2}) tends to
zero: $\{\delta,\epsilon,\mu,\nu\} \rightarrow 0$. In particular we will focus on
the solution marked with a purple circle on the {\bf C} family in the inset of
Fig.\ref{fig1}. We will compute its stability region when the gain, loss and
higher order Kerr terms are gradually suppressed in the Ginzburg-Landau model. 

\begin{figure}
\centering
\begin{tabular}{cc}
\epsfig{file=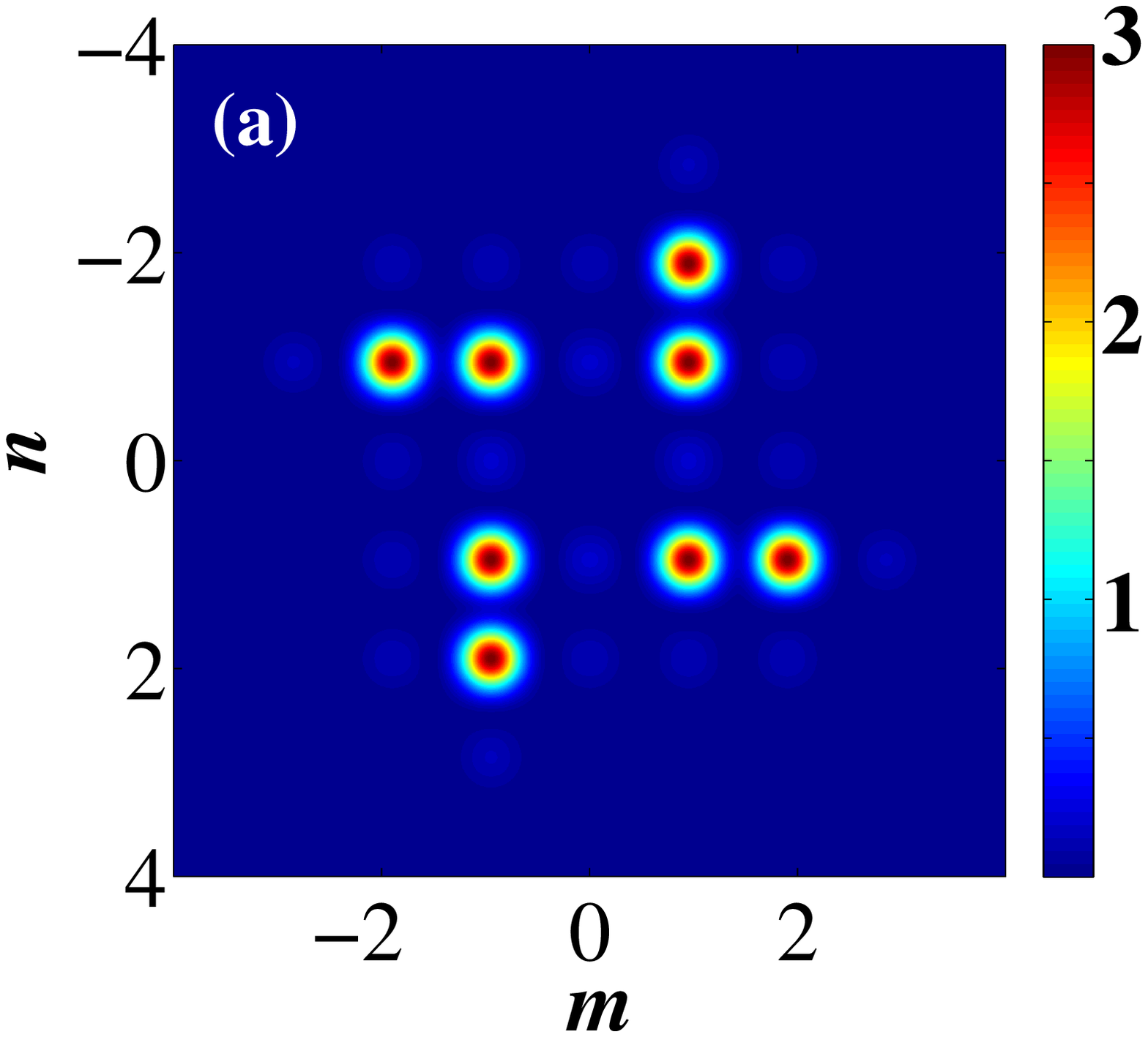,width=0.48\linewidth,clip=}&
\epsfig{file=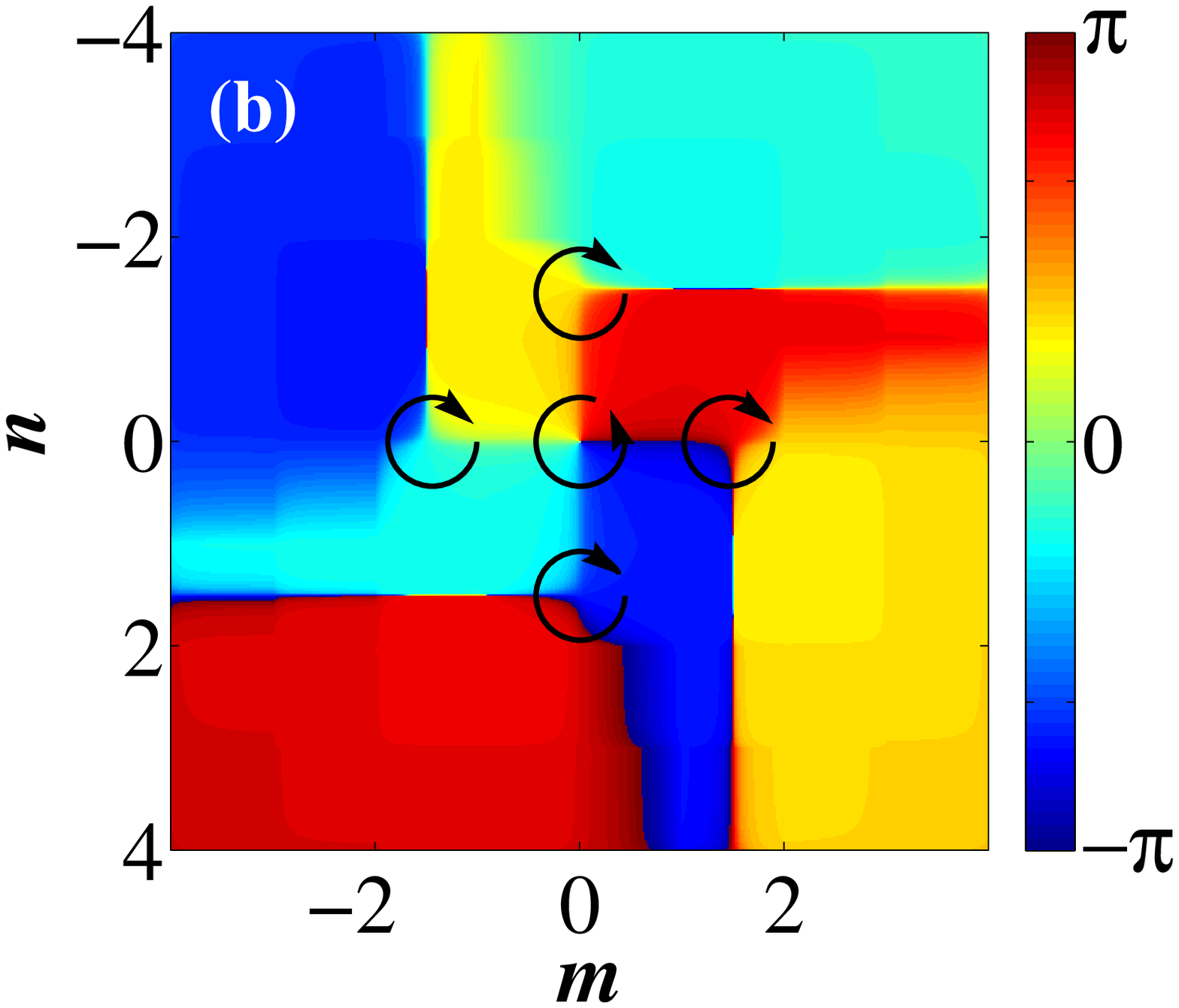,width=0.47\linewidth,clip=}
\end{tabular}
\caption{(Color online) Color map plots for the eight peaks stable two charges 
($S=1$ and $S=-3$) vortex solution localized on the {\bf C} family at the purple 
circle in the Fig.\ref{fig1}. (a) Amplitude profile. (b) Phase profile.}
\label{fig9}
\end{figure}

\begin{figure}
\centering
\epsfig{file=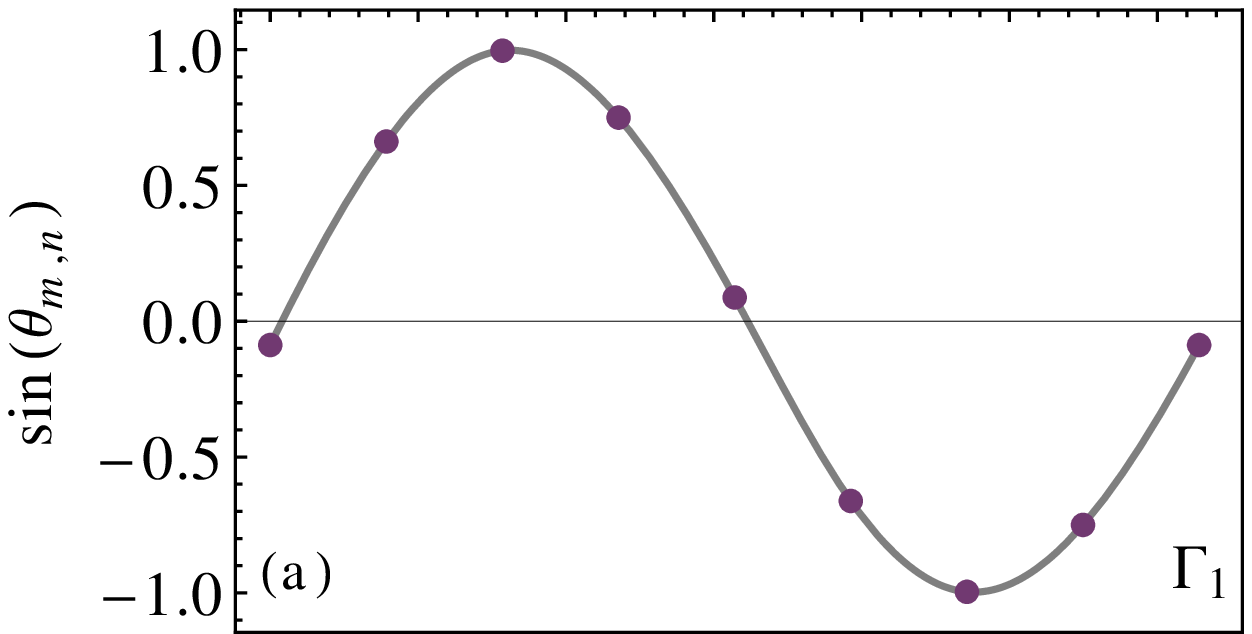,width=0.75\linewidth,clip=}\\
\epsfig{file=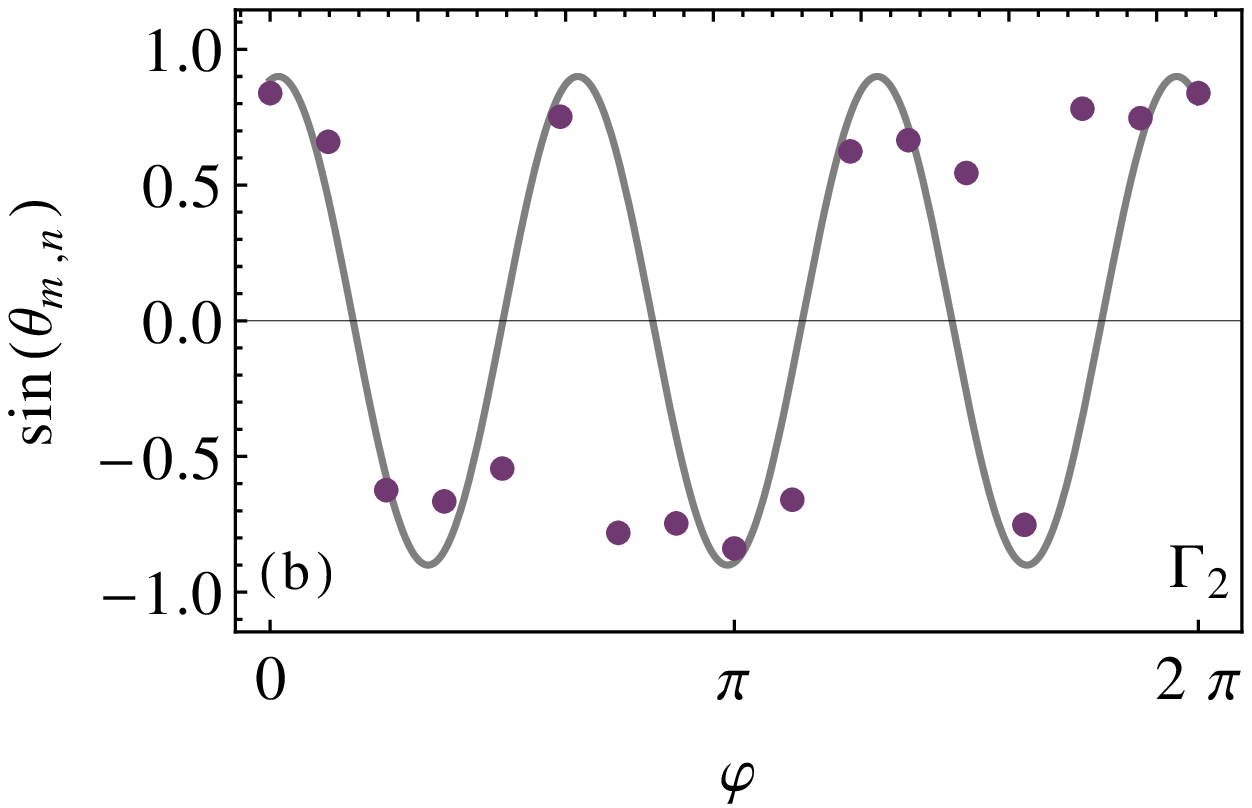,width=0.75\linewidth,clip=}
\caption{(Color online) $\sin(\theta_{m,n})$ versus $\varphi$ (azimuthal
angle for the lattice) diagram for the first (a) and second (b) 
discrete contour for the swirl-vortex soliton, marked with a purple 
dot on the {\bf C} family in Fig.\ref{fig1}.}
\label{fig10}
\end{figure}

These solutions are of the ``two-charges" vortex type, with charges $S=1$ and
$S=-3$~\cite{PhysRevA.82.063818}. Moreover, this type of solutions (the
swirl-vortex soliton) can be understood as a bound state of five
vortices~\cite{PhysRevA.79.053820,Chong2009126}. Indeed, we can identify a
vortex with $S=1$ at the origin ({\large $\circlearrowleft$} symbol), surrounded
by four vortex, each with $S=-1$, whose singularities are located at the center
of the {\large $\circlearrowright$} symbols on the Fig.\ref{fig9}(b). This
interpretation agrees with the transition of the effective vorticity from
$S=1\rightarrow S=-3$, as we move farther from the center. The amplitude and
phase profiles for this solution are displayed in Fig.\ref{fig9}(a-b). It should
be noted that it has the same power value that the solution marked with the
green circle on the {\bf A} family. (In fact, we plot its corresponding family
in the inset because both families have almost identical $Q$ versus
$\varepsilon$ diagrams).
We can see that both amplitude profiles are very different; even though each one
has eight principal excited sites, their spatial distributions are dissimilar.
In addition, their phase profiles are completely different. While the solutions
on the {\bf A} family have a well defined unique topological charge $S=3$, the
solutions on the {\bf C} family have two charges ($S=1$ and $S=-3$) simultaneously,
as mentioned before. Again, we plot $\sin(\theta_{m,n})$ vs $\varphi$ for the
first ($\Gamma_{1}$) and the second ($\Gamma_{2}$) discrete contour. From
Fig.\ref{fig10}(a) we can see one period ($S=1$)  for the sinusoidal function
(gray line) along the first contour, and for the second contour we have three
periods ($S=-3$) as shown in Fig.\ref{fig10}(b).
Unlike the conservative cubic case (NLSE), in the dissipative model the
propagation constant $\lambda$ is not an arbitrary parameter that can be chosen
at will. It is fixed by the rest of the CQGL equation parameters. By changing
them, the value of the propagation constant also changes.  As in other nonlinear
problems~\cite{Soto-Crespo:00}, we can think of the dissipative terms as
determinant to select one of the infinite solutions of the associated
conservative problem. With this in mind we will find out the stability
regions in terms of the propagation constant so we can compare with the
Schr\"odinger limit. 

\begin{figure}[htbp]
\centering
\begin{tabular}{cc}
\epsfig{file=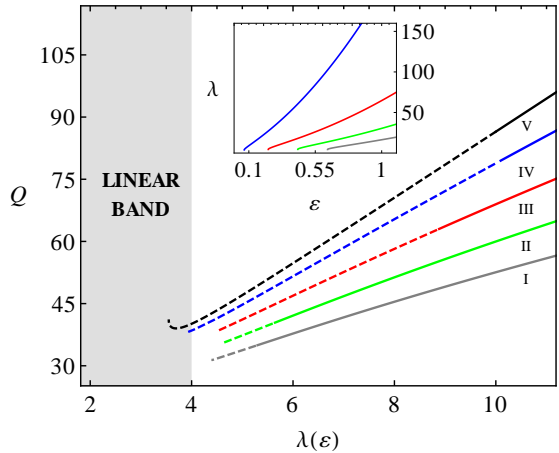,width=0.85\linewidth,clip=}
\end{tabular}
\caption{(Color online) $Q$ versus $\lambda(\varepsilon)$ diagram for several
sets of parameters specified in Table I, of two charges swirl-vortex
solitons. Inset shows $\lambda$ vs $\varepsilon$.}
\label{fig11}
\end{figure}

For the sake of comparison we construct the $Q$ versus $\lambda$ diagram shown
in Fig.\ref{fig11}. Here, we have fixed $\delta$, $\mu$ and $\nu$ parameters and
we only move through the $\varepsilon$ parameter (nonlinear gain). In this way,
we obtain a solution and its corresponding propagation constant for each value
of $\varepsilon$. Then, we proceed varying the rest of the parameters slightly,
and construct a new curve, taking the solutions of the previous curve as 
initial conditions in our multidimensional Newton-Raphson scheme. In the inset of
Fig.\ref{fig11} we show the corresponding $\lambda$ vs $\varepsilon$ diagram.

With the previous scheme we can find a large number of curves, but for the
sake of clarity, we only show three of them; they are located between the
conservative cubic case (black branch) and the {\bf C} curve (gray branch). We
can read from Table~\ref{disspar} the CQGL equation parameters corresponding to
the curves displayed in Fig.\ref{fig11}. These five branches belong to the same
family conformed by vortex solutions with amplitude and phase profiles such 
as those showed in Fig.\ref{fig9}.

We have done the standard linear stability analysis, described in section II,
for each one of them. The part of the curves that correspond to  stable
solutions are shown with continuous lines while the dashed lines correspond to
unstable solutions.

Taking the above into account, we can clearly  establish that if the dissipation
is attenuated the stability regions for the soliton solutions are reduced.
\cite{Soto-Crespo:00} Indeed, we can see here a wide difference between the
stability regions for the Schr\"odinger limit and the {\bf C} branch. The first
one only has stable solutions for propagation constant values far away from the
linear band, the last one has stable solutions for propagation constant values
closer to the linear band.

\begin{table}[ht]
\caption{CQGL equation parameters} 
\centering  
\begin{tabular}{c c c c} 
\hline\hline\\ [0.001ex]
Curve &\hspace{1cm}$\delta$\hspace{1cm}&\hspace{1cm}$\mu$\hspace{1cm}&\hspace{1cm}$\nu$\hspace{1cm}\\ [0.2ex] 
\hline\\ [0.01ex]    
{\scriptsize I} &\hspace{1cm}-0.9\hspace{1cm}&\hspace{1cm}-0.1\hspace{1cm}&\hspace{1cm}0.1\hspace{1cm}\\
{\scriptsize II} &\hspace{1cm}-0.8\hspace{1cm}&\hspace{1cm}-0.08\hspace{1cm}&\hspace{1cm}0.08\hspace{1cm}\\
{\scriptsize III} &\hspace{1cm}-0.4\hspace{1cm}&\hspace{1cm}-0.03\hspace{1cm}&\hspace{1cm}0.03\hspace{1cm}\\
{\scriptsize IV} &\hspace{1cm}-0.1\hspace{1cm}&\hspace{1cm}-0.01\hspace{1cm}&\hspace{1cm}0.01\hspace{1cm}\\
{\scriptsize V} &\hspace{1cm}0\hspace{1cm}&\hspace{1cm}0\hspace{1cm}&\hspace{1cm}0\hspace{1cm}\\ [0.7ex]      
\hline 
\hline 
\end{tabular}
\label{disspar} 
\end{table}

\section{Summary and conclusions}
\label{con}
In conclusion, we have found discrete vortex solitons (symmetric and asymmetric)
with higher-order vorticity in dissipative 2D-lattices and studied its
stability. In particular, we have also shown in detail a solution that contains
two topological charges. Finally, we analyzed the stability of the solutions
when dissipation in the system is decreased, observing that the stability
regions shrink. A comparison with the conservative cubic case is done, showing
that dissipation serves to provide stability to otherwise unstable conservative
solutions.

\section{Acknowledgments}

C.M.C. and J.M.S.C. acknowledge support from the Ministerio de Ciencia e
Innovaci\'on under contracts FIS2006-03376 and FIS2009-09895. R.A.V and M.I.M
acknowledge support from FONDECYT, Grants 1080374 and 1070897, and from 
Programa de Financiamiento Basal de CONICYT (FB0824/2008).

\bibliography{../dvortex}

\end{document}